# Fe-doping induced superconductivity in charge-density-wave system $1T$-TaS$_2$

(Also see: Euro. Phys. Lett. 97, 67005 (2012).)

L. J. Li[1], W. J. Lu[1] [(a)], X. D. Zhu[2], L. S. Ling[2], Z. Qu[2] and Y. P. Sun[1,2] [(b)]

[1] *Key Laboratory of Materials Physics, Institute of Solid State Physics, Chinese Academy of Sciences, Hefei 230031, People's Republic of China*
[2] *High Magnetic Field Laboratory, Chinese Academy of Sciences, Hefei 230031, People's Republic of China*

**Abstract** – We report the interplay between charge-density-wave (CDW) and superconductivity of $1T$-Fe$_x$Ta$_{1-x}$S$_2$ ($0 \leq x \leq 0.05$) single crystals. The CDW order is gradually suppressed by Fe-doping, accompanied by the disappearance of pseudogap/Mott-gap as shown by the density functional theory (DFT) calculations. The superconducting state develops at low temperatures within the CDW state for the samples with the moderate doping levels. The superconductivity strongly depends on $x$ within a narrow range, and the maximum superconducting transition temperature is 2.8 K as $x = 0.02$. For high doping level ($x > 0.04$), the Anderson localization (AL) state appears, resulting in a large increase of resistivity. We present a complete electronic phase diagram of $1T$-Fe$_x$Ta$_{1-x}$S$_2$ system that shows a dome-like $T_c(x)$.

**Introduction.** – The relationship between superconductivity and charge-density-wave (CDW) is debated for several decades. The idea that superconductivity and CDW can be competing and even coexisting electronic states is of fundamental interest in condensed matter physics [1–6]. For hole-doped high-$T_c$ cuprates La$_{2-x}$Sr$_x$CuO$_4$, the magnetic order and superconductivity has almost no overlapping from the phase diagram [7], while for electron-doped high-$T_c$ cuprates Nd$_{2-x}$Ce$_x$CuO$_4$, the overlap is clearly present [7,8]. For Fe-based superconductor the coexistence of spin-density-wave (SDW) and superconductivity has been clearly observed [1–4,9]. In transition-metal dichalcogenide compounds (TMDCs) with the form of MX$_2$ (where M is a group 4 or 5 metal and X=S, Se, or Te), for the host compound $2H$-NbSe$_2$, Kiss *et al.* reported that the CDW order can boost the superconductivity [10], while Borisenko *et al.* reported that the CDW order forestalls the superconducting gap by excluding the nested portions of the Fermi surface from participating in superconductivity. For the intercalated compound $1T$-Cu$_x$TiSe$_2$, the CDW and superconductivity states can coexist [12–14], while there is a competition between the superconductivity and the CDW in $2H$-Na$_x$TaS$_2$ system [15]. So the nature of interplay between CDW and superconductivity deserves further investigation, and the layered TMDCs may be the ideal compounds for this

[(a)]Corresponding author. E-mail: wjlu@issp.ac.cn
[(b)]Corresponding author. E-mail: ypsun@issp.ac.cn

L. J. Li, W. J. Lu, X. D. Zhu, L. S. Ling, Z. Qu, and Y. P. Sun

program.

In the family of TMDCs, two basic structures of $TaS_2$ were found and defined by the different orientation of stacking chalcogensheets. One is $2H$-$TaS_2$ with Ta in trigonal prismatic coordination with S atoms, and the other one is $1T$-$TaS_2$ with Ta in octahedral coordination with S atoms. For $1T$-$TaS_2$, with decreasing temperature, it exhibits incommensurate, nearly commensurate, and commensurate CDWs [16,17]. The commensurate CDW (CCDW) phase accompanies periodic lattice distortions that form a supercell with star-of-David clusters [18], consisting of two 6-atom rings which contract towards the central atom. The CDW is accompanied by a period lattice distortion and hence induces the Peierls gap. Earlier empirical tight-binding calculations of the band structure predicate a narrow half-filled band straddling the Fermi level and the Peierls gap located at -50 meV below the Fermi level [19]. However some experimental and theoretical works show the gap is due to the electron correlation effect, *i.e.* Mott localization gap [20–22]. In real space, the Mott transition corresponds to the localization of one Ta-$5d$ electron at the center of the David-star cluster unit of the CDW. In reciprocal space, it is due to the opening of a correlation gap in a narrow half-filled band straddling the Fermi level.

The CDW insulating phase in $1T$-$TaS_2$ system can be inhibited by the external pressure [17], disorder [23], and photo-excitation [24], and hence the metallic state or superconductivity may arise in this system. It is worthwhile to investigate $1T$-$TaS_2$ system by doping, since the doping can effectively modify the band structure by altering the crystal structure and carriers concentration, and thereby affect the CDW state and potentially induce new collective states. In this paper, we present the combined experimental and theoretical studies of the superconductivity of $1T$-$Fe_xTa_{1-x}S_2$ single crystals and report a complete electronic phase diagram of $1T$-$Fe_xTa_{1-x}S_2$ system which shows a dome-like $T_c(x)$. This finding complements the recent discovery of pressure induced superconductivity in $1T$-$TaS_2$ [17].

**Experiment and theoretical details.** – The $1T$-$Fe_xTa_{1-x}S_2$ single crystals were grown via the chemical vapor transport (CVT) method with iodine as a transport agent, and the sample stoichiometry was verified by X-ray energy dispersive spectroscopy (EDS) of a scanning electron microscope (SEM) and inductively coupled plasma atomic emission spectrometry (ICP-AES). Taking account of the possible inhomogeneity existing in the samples, we selected multiple points on crystal surface and analyzed. The measured results show that the actual concentration $x$ is close to the nominal one. The resistivity and the Seebeck effect were measured in a Quantum Design physical property measurement system (PPMS), and the magnetization measurements were performed on a superconducting quantum interference device (SQUID) system. The electronic structure calculations were performed in the framework of density functional theory (DFT) within the local-density approximation (LDA), using the ABINIT code [25]. Normconserving pseudopotentials and a plane-wave basis set with a kinetic energy cutoff of 1600 eV were used. For the undistorted $1T$-phase a $14 \times 14 \times 8$ Monkhorst $k$-point sampling was used, while for the CCDW phase, a $6 \times 6 \times 8$ mesh of $k$-point was used. In order to describe the strong correlation of electrons in Mott-Hubbard physics, we adopted a LDA+$U$ method where $U$ is the onsite Coulomb repulsion. For the Fe-doped system, we model the $\sqrt{13} \times \sqrt{13} \times 2$ supercell which contains at least one-site substitution of Ta atom by Fe atom (1/26 doping level). The $\sqrt{13} \times \sqrt{13} \times 2$ supercell corresponds to the CCDW in the basal Ta-plane with the assumed simple stacking out of plane.

**Results and discussions.** – Our x-ray diffraction (XRD) measurement results (not shown in figures) show the lattice constant $a$ decreases with an increase in $x$, while the lattice constant $c$ almost has no change. This is different from the case of the intercalated systems. The observed reduction of lattice constant should be due to the substitution of Fe at Ta site, since the ion radius of Fe is smaller than that of Ta. The temperature dependence of in-plane resistivity ($\rho_{ab}$) measurements on $1T$-$Fe_xTa_{1-x}S_2$ single crystals using standard 4-point technique are shown in Fig. 1(a). For the $x \leq 0.04$ samples, the resistivity decreases with the increasing of $x$ until $x > 0.02$ and then increases. Especially a large resistivity increase in the low-temperature range appears for the $x = 0.05$ sample. The anomalous resistivity increase at low temperatures was previously observed for $1T$-Fe$_x$Ta$_{1-x}$S$_2$ with high Fe-doping level ($x \geq 0.05$), which has been attributed to the Anderson localization (AL) [26]. The Ta-site disorder can be induced by the random atom occupation due to the Fe doping, which produces a random potential and localizes the electrons [26–28]. For our $x = 0.05$ sample the AL insulating state potentially exists at low temperatures. Furthermore, the low-temperature CCDW phase is suppressed for $x \geq 0.01$, and interestingly the signature of superconductivity appears at 2.8 and 2.6 K for $x = 0.02$ and 0.03, respectively, while no superconductivity appears for $x \geq 0.04$ down to 2 K (see the inset of Fig. 1(a)). The maximum zero resistivity temperature is about 2.1 K at $x = 0.02$. So it may be the optimal doping and we study this composition in details in the following parts.

Figure 1(b) shows the high temperature resistivity of $1T$-Fe$_x$Ta$_{1-x}$S$_2$ single crystals. The samples show sudden increases of resistivity corresponding to the nearly commensurate CDW (NCCDW) transitions for $x \leq 0.04$. The inset of Fig. 1(b) shows the temperature dependence of thermopower ($S$) for the $x = 0$ and 0.03 samples. For the $x = 0$ sample, $S$ changes signs at about 350 K and 140 K, which can be attributed to the NCCDW and CCDW transitions, respectively [29]; while there is only one transition at about 340 K for the $x = 0.03$ sample, in agreement with the resistivity result. This may indicate that the CCDW phase disappears, while the NCCDW phase still remains even after 3% Fe doping. Moreover, the Seebeck coefficient of the $x = 0.03$ sample becomes more negative in the whole temperature range, suggesting that Fe acts at least in part as an $n$-type doping.

Magnetic characterization of the superconducting transition for $1T$-Fe$_{0.02}$Ta$_{0.98}$S$_2$ at 10 Oe as $H \parallel ab$ is shown in Fig. 2. The diamagnetism in the low temperature region further confirms the existence of superconductivity, and the steep transition in $M(T)$ curve indicates that the sample is rather homogeneous. The superconducting transition temperature is about 2.1 K that is defined by the onset point of the zero-field-cooling (ZFC) and field-cooling (FC) curves . The smaller magnetization value for FC is likely due to the complicated magnetic flux pinning effects [30]. The value of $-4\pi\chi$ at 0.5 K is about 75% where the demagnetization factor is negligible for $H \parallel ab$, implying the bulk superconductivity of $1T$-Fe$_{0.02}$Ta$_{0.98}$S$_2$ sample.

The right top inset of Fig. 2 shows the initial $M(H)$ curve of $1T$-Fe$_{0.02}$Ta$_{0.98}$S$_2$ single crystal in the low field region at 0.75 K as $H \parallel ab$, which allows us to estimate the lower critical field values ($H_{c1}$) at 0.75 K. At low fields, the $M(H)$ isotherm is linear in $H$, as expected for a BCS type-II superconductor. We estimate the $H_{c1}(0.75K)$ value to be about 14 Oe, marked by arrow, from the point where this curve deviates from linearity. The obtained slope of the linear fitting up to 15 Oe of our experimental data at 0.75 K is -0.98(6), which corresponds to $-4\pi M = H$ as $H \parallel ab$, describing the Meissner shielding effects. The right bottom inset of Fig. 2 shows the magnetization hysteresis loop of $1T$-Fe$_{0.02}$Ta$_{0.98}$S$_2$ at 0.5 K as $H \parallel ab$, and the shape of the $M(H)$ curve further indicates that $1T$-Fe$_{0.02}$Ta$_{0.98}$S$_2$ is a typical type-II superconductor.

In order to elucidate the Fe-doping effect on the properties of the host $1T$-TaS$_2$, we carried out theoretical calculation by DFT. Figures 3(a) and 3(b) show the calculated Fermi surfaces for the undistorted $1T$ structure and the reconstructed CCDW superstructure, respectively, which clearly show that the undistorted $1T$-TaS$_2$ has a quasi-two-dimensional Fermi surface, while the CCDW phase has a quasi-one-dimensional Fermi surface. Recent theoretical calculations show that the CCDW reconstruction induces drastic change in the electronic structure with a pseudogaped Fermi suface [31,32]. The angle-resolved photoemission spectroscopy (ARPES) measurement shows the Fermi surface nesting feature and the CDW distortion enhanced electron-phonon coupling in $1T$-TaS$_2$ [33]. Figure 3(c) shows the calculated energy band structure near the Fermi level $E_F$ for the ideal/undistorted $1T$-phase and the reconstructed CCDW superstructure. For the ideal/undistorted $1T$ structure, the Ta-$d$ bands cross the Fermi level and strongly disperse along $\Gamma$-$M$-$K$-$\Gamma$ and $A$-$L$-$H$-$A$, which leads to a good metallic character. Along $\Gamma$-$A$ only weak dispersion on the Fermi level is left, consistent with the quasi-2D character of the 1$T$-phase structure. The relaxed CCDW structure shows the star-of-David clusters in which the first and second rings of Ta atoms averagely contract inwards by 5-6%. The Ta star-of-David clusters lead the Ta-$d$ state to become localized in the in-plane directions, which results in a localized uppermost band along $\Gamma$-$M$-$K$-$\Gamma$ at about -0.3 eV. The band at the Fermi level disperses only along $\Gamma$-$A$ direction, indicating the existence of quasi-one-dimensional Fermi surface (see Fig. 3(b)) that allows that electrons conduct only along the $c$-axis, and the $ab$-plane conductivity becomes worse in CCDW structure. Recently Rossnagel *et al.* [34] suggest that in addition to the CDW distortion, the spin-orbit effect could have a large influence on the bands near the Fermi level in 1$T$-TaS$_2$. By considering the effects of the spin-orbit interaction, Freericks *et al.* [20] calculated the band structure of CCDW-ordered 1$T$-TaS$_2$ and found that the spin-orbit interaction splits off the center Ta-site nonbonding band from higher-energy bands and the band still has a width of about 0.4 eV due to $k_z$ ($\Gamma$-$A$) dispersion.

Figure 3(d) shows the density of states (DOS) for the undistorted 1$T$-TaS$_2$ and the distorted CCDW structure. For the undistorted 1$T$-phase, a large DOS locates at the Fermi level. However, in the CCDW structure, the DOS near the Fermi level shows a pseudogap structure, resulting in a semimetal conduction. The exprimentally observed gap from optical conductivity and ARPES measurements is a real gap on the order of 0.12-0.18 eV [35, 36], which was attributed to the Mott localization gap arising from the strong correlation of Ta-$d$ electrons [21, 22]. In order to describe the Mott-Hubbard physics, we applied LDA+$U$ method to the CCDW structure, where the effective on-site Coulomb $U$ is set to be 0.5 eV (by using this value, the calculated gap is closed to the experimentally observed gap of 180 meV from ARPES [36]). The calculated results show that a deep Mott gap develops and the localized states form at the Fermi level, which accounts for the experimentally observed insulating conduction behavior at low temperatures. We note that Freericks *et al.* [20] used a dynamical mean-field theory (DMFT) approach to calculate the local DOS in which a gap on the order of 0.125-0.15 eV was produced. This gap is slightly smaller than our obtained one. They also pointed out the upper and lower Hubbard bands were merging with the next higher and next lower bands in the band structure, indicating that the single-band model may be inadequate for this system. By Fe-doping, the energy gap disappears due to the suppression of CCDW distortions (see Fig. 3(e)), and also the Fe-3$d$ band partially contributes to the DOS at the Fermi level (the calculated atom-projected DOS are not shown in figures), which could contribute to the metallic character for Fe$_x$Ta$_{1-x}$S$_2$ and consistent with the thermopower results.

Finally, the overall behavior of this system is summarized as the electronic phase diagram presented in Fig. 4. By using Fe doping as a controlled tuning parameter, the CDW transition of 1$T$-TaS$_2$ is driven in temperature, and a new superconducting state emerges. The dependence of the superconducting transition temperature on the iron content $T_c(x)$ shows a domelike structure, characteristically found in the electronic phase diagrams of some heavy fermion and layered organic superconductors [37–39]. Furthermore, the superconductivity strongly depends on $x$, and only occurs in the narrow doping level. The superconductivity appears for $x > 0.01$, going through a maximum of 2.8 K as $x = 0.02$, followed by a decrease before the chemical phase boundary ($x = 0.04$) is reached. The long-range CCDW phase is suppressed by Fe-doping and develops into short-range NCCDW phase and metallic interdomain regions. The induced superconductivity and the CDW may compete with each other and probably are separated in real space [17]. On the other hand the electron-phonon interaction in CDW domains may also play an important role in the observed superconductivity [40]. The similar superconductivity induced by K-doping upon inhibiting the long-range CDW order in BaBiO$_3$ system was observed previously [41]. As $x \geq 0.05$, the Anderson localization state due to the random potential, which loses the conduction-electron mobility, results in the resistivity increase at low temperatures.

**Conclusion.** – In summary, we have systematically investigated the $1T$-Fe$_x$Ta$_{1-x}$S$_2$ single crystals by experiments and theoretical calculations and presented the electronic phase diagram for $1T$-Fe$_x$Ta$_{1-x}$S$_2$ system. We found that with the suppression of the CCDW, the superconducting state appears at low temperatures as $0.01 < x < 0.04$, while the NCCDW exists all the same. It may suggest that these two phases can coexist and keep the balance in favor of superconductivity until $x \geq 0.04$. The maximum superconducting transition temperature is 2.8 K as $x = 0.02$, and a dome-like $T_c(x)$ is obtained. The calculated DOS near the Fermi level for the undoped and the Fe-doped $1T$-TaS$_2$ show that the Mott gap disappears after Fe doping and the Fe-$3d$ band partially contributes to the DOS at the Fermi level, which may contribute to the metallic and superconducting character. As $x > 0.04$, the Anderson localization state appears, resulting in a large increase of resistivity at low temperatures.

∗ ∗ ∗

This work was supported by the National Key Basic Research under contract No. 2011CBA00111, and the National Nature Science Foundation of China under contract Nos. 10804111, 10974205, 11104279 and 51102240, and Director's Fund of Hefei Institutes of Physical Science, Chinese Academy of Sciences. The DFT calculations were partially performed at the Center for Computational Science of CASHIPS.

L. J. Li, W. J. Lu, X. D. Zhu, L. S. Ling, Z. Qu, and Y. P. Sun

**Figure captions**

**FIG. 1.** (a) Temperature dependent in-plane resistivity $\rho_{ab}(T)$ of $1T$-Fe$_x$Ta$_{1-x}$S$_2$ single crystals. Inset shows the detail of superconducting transitions. (b) The NCCDW transitions near 350 K of all the samples. Inset shows the temperature dependence of the thermopower of $1T$-Fe$_x$Ta$_{1-x}$S$_2$ crystals for $x = 0$ and 0.03.

**FIG. 2.** Temperature dependence of magnetic susceptibility of $1T$-Fe$_{0.02}$Ta$_{0.98}$S$_2$ single crystals. The right top inset shows the initial $4\pi M(H)$ isotherm at 0.75 K, and the red line shows the linear fitting in the low field range. The right bottom inset shows the magnetization hysteresis loops of $1T$-Fe$_{0.02}$Ta$_{0.98}$S$_2$ at $T = 0.5$ K as $H \parallel ab$.

**FIG. 3.** (a) Fermi surface for the undistorted $1T$ structure and (b) reconstructed C-CDW superstructure. (c) Energy band structure near the Fermi level for the undistorted $1T$-phase (blue points) and the reconstructed CCDW superstructure (solid lines). The band that crosses the Fermi level is plotted as a red line. (d) Density of states near the Fermi level for the un-doped $1T$-TaS$_2$ and (e) the Fe-doped system.

**FIG. 4.** The electronic phase diagram of $1T$-Fe$_x$Ta$_{1-x}$S$_2$ single crystals.

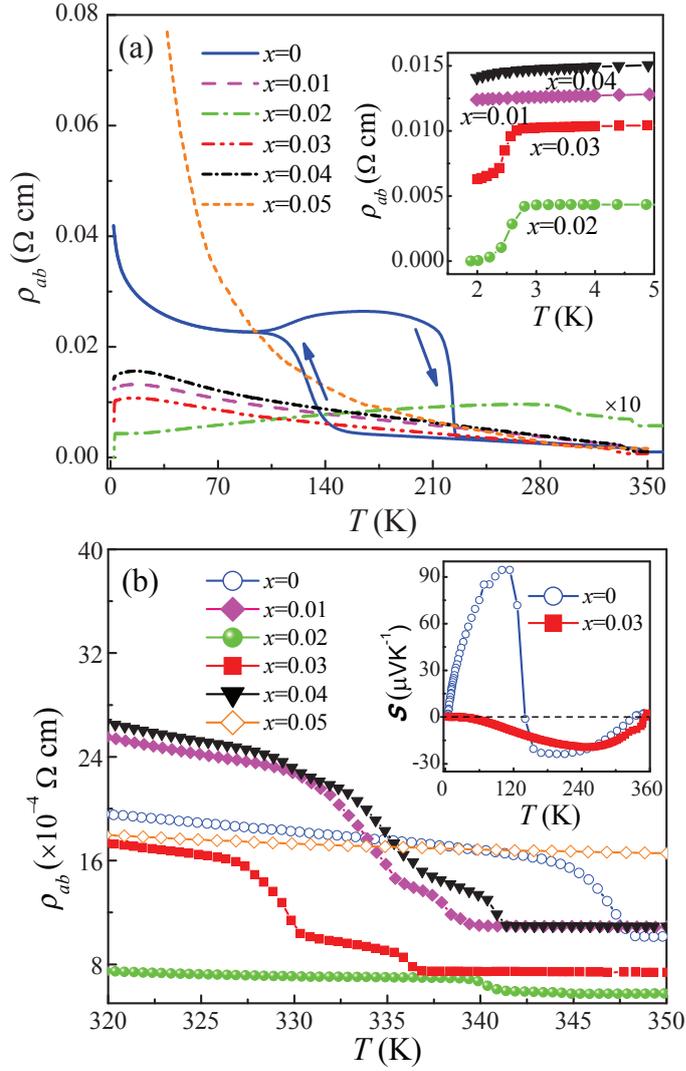

Fig. 1: (a) Temperature dependent in-plane resistivity $\rho_{ab}(T)$ of 1T-Fe$_x$Ta$_{1-x}$S$_2$ single crystals. Inset shows the detail of superconducting transitions. (b) The NCCDW transitions near 350K of all the samples. Inset shows the temperature dependence of the thermopower of 1T-Fe$_x$Ta$_{1-x}$S$_2$ crystals for $x = 0$ and 0.03.

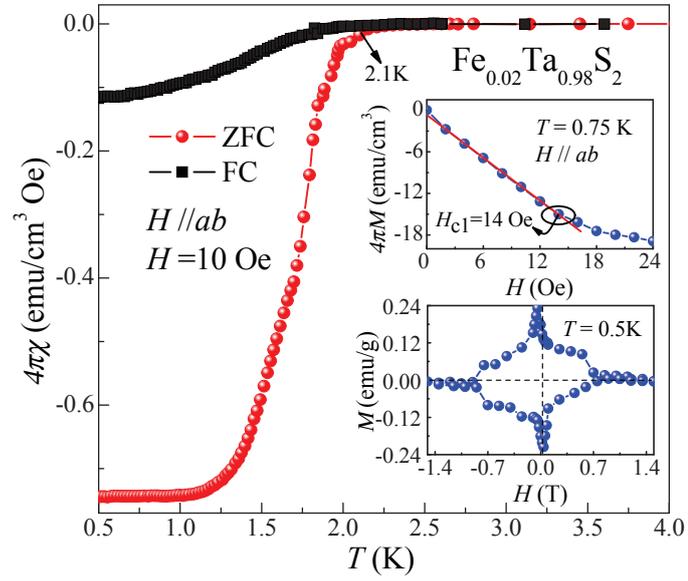

Fig. 2: Temperature dependence of magnetic susceptibility of 1$T$-Fe$_{0.02}$Ta$_{0.98}$S$_2$ single crystals. The right top inset shows the initial $4\pi M(H)$ isotherm at 0.75 K, and the red line shows the linear fitting in the low field range. The right bottom inset shows the magnetization hysteresis loops of 1$T$-Fe$_{0.02}$Ta$_{0.98}$S$_2$ at $T = 0.5$ K as $H \parallel ab$.

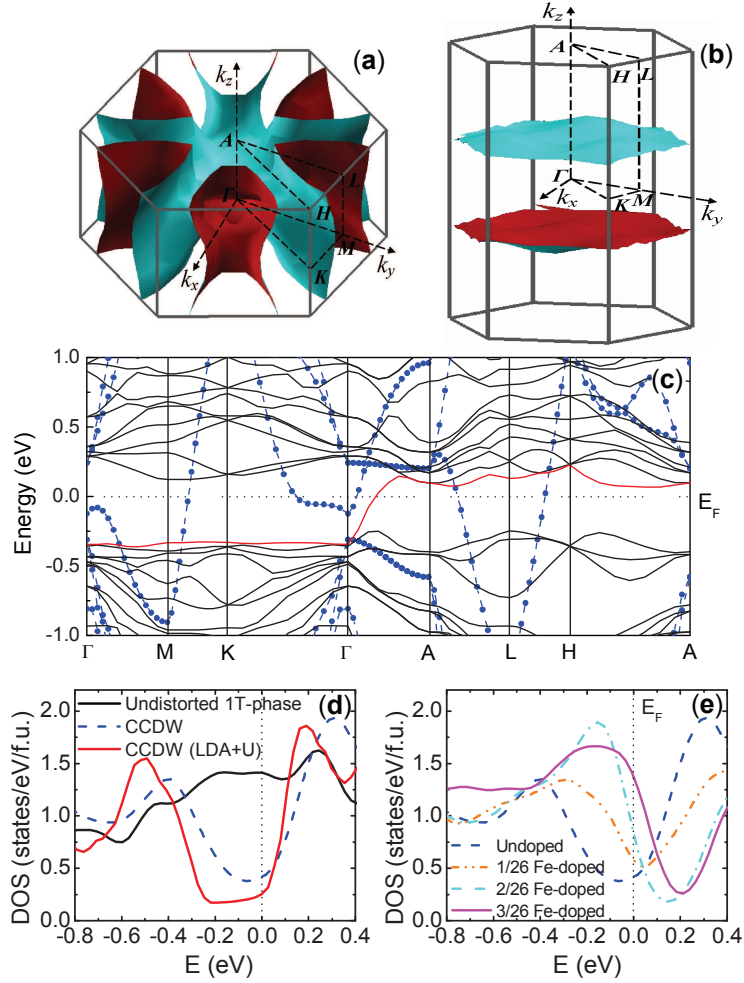

Fig. 3: (a) Fermi surface for the undistorted $1T$ structure and (b) reconstructed CCDW superstructure. (c) Energy band structure near the Fermi level for the undistorted $1T$-phase (blue points) and the reconstructed CCDW superstructure (solid lines). The band that crosses the Fermi level is plotted as a red line. (d) Density of states near the Fermi level for the un-doped $1T$-TaS$_2$ and (e) the Fe-doped system.

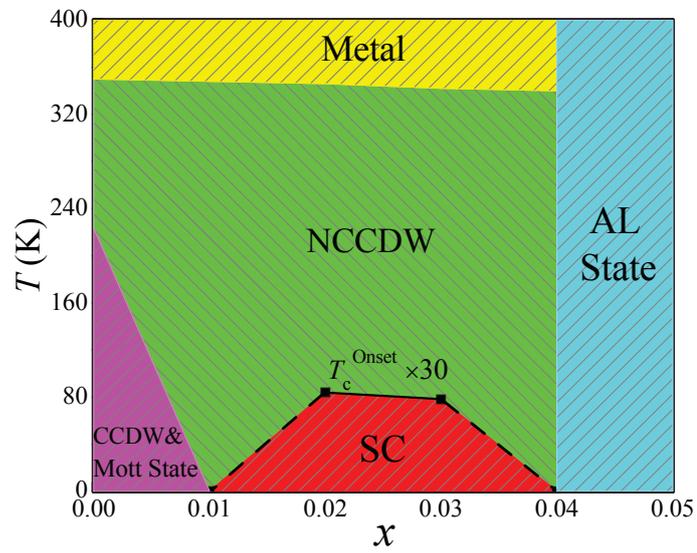

Fig. 4: The electronic phase diagram of $1T$-$Fe_x Ta_{1-x} S_2$ single crystals.